\documentclass[nofootinbib, reprint, preprintnumbers, amsmath, amssymb, amsfonts, aps, pra, superscriptaddress, floatfix]{revtex4-2}
\usepackage[utf8]{inputenc}
\usepackage{mathtools}
\usepackage{graphicx}
\usepackage{dcolumn}
\usepackage{bm}
\usepackage{physics}
\usepackage{color}
\usepackage{braket}
\usepackage[normalem]{ulem}
\usepackage{enumerate}
\usepackage{enumitem}
\usepackage[colorlinks=true,linkcolor=blue,citecolor=blue,urlcolor =blue]{hyperref}
\usepackage{mathrsfs}
\usepackage{bbm}
\usepackage{siunitx}
\usepackage{graphicx,color}
\usepackage{amssymb}
\usepackage{amsmath}
\usepackage{bm}
\usepackage{comment}
\usepackage{lipsum}
\newcommand{\YM}[1]{\textcolor[rgb]{0, 0, 0}{#1}}

\newcommand{\AdMORI}[1]{\textcolor[rgb]{0, 0, 0}{#1}}

\begin{document}
\title{
Single-qubit rotations on a binomial code
without ancillary qubits
}

\author{Yuki Tanaka}
\altaffiliation{These authors equally contributed to this paper.}
\affiliation{Department of Electrical, Electronic, and Communication Engineering, Faculty of Science and Engineering, Chuo university, 1-13-27, Kasuga, Bunkyo-ku, Tokyo 112-8551, Japan}%
\affiliation{NEC-AIST Quantum Technology Cooperative Research Laboratory, National Institute of Advanced Industrial Science and Technology (AIST), Tsukuba, Ibaraki 305-8568, Japan}

\author{Yuichiro Mori}
\altaffiliation{These authors equally contributed to this paper.}
\email{mori-yuichiro.9302@aist.go.jp}
\affiliation{Global Research and Development Center for Business by Quantum-AI Technology (G-QuAT), National Institute of Advanced Industrial Science and Technology (AIST), 1-1-1, Umezono, Tsukuba, Ibaraki 305-8568, Japan}%

\author{Yuta Shingu}
\affiliation{Department of Physics, Graduate School of Science, Tokyo University of Science, Shinjuku, Tokyo 162-8601, Japan.}

 \author{\\Aiko Yamaguchi}
 \affiliation{NEC-AIST Quantum Technology Cooperative Research Laboratory, National Institute of Advanced Industrial Science and Technology (AIST), Tsukuba, Ibaraki 305-8568, Japan}
 \affiliation{Secure System Platform Research Laboratories, NEC Corporation, 
1753, Shimonumabe, Kawasaki, Kanagawa 211-0011, Japan}

\author{Tsuyoshi Yamamoto}
\affiliation{NEC-AIST Quantum Technology Cooperative Research Laboratory, National Institute of Advanced Industrial Science and Technology (AIST), Tsukuba, Ibaraki 305-8568, Japan}
\affiliation{Secure System Platform Research Laboratories, NEC Corporation, 
1753, Shimonumabe, Kawasaki, Kanagawa 211-0011, Japan}

\author{Yuichiro Matsuzaki}
\email{ymatsuzaki872@g.chuo-u.ac.jp}
\affiliation{Department of Electrical, Electronic, and Communication Engineering, Faculty of Science and Engineering, Chuo university, 1-13-27, Kasuga, Bunkyo-ku, Tokyo 112-8551, Japan}%

\date{\today}


\begin{abstract}
Great attention has been paid to binomial codes utilizing bosonic systems as logical qubits with error correction capabilities. However, implementing single-qubit rotation operations on binomial codes has proven challenging, requiring an ancillary qubit in previous approaches. Here, we propose a method for performing logical qubit rotation on binomial codes without requiring an 
\YM{ancillary} qubit. Specifically, we explain how to implement $X$-axis rotations by simultaneously applying two-frequency parametric drives to resonators with nonlinearity. Furthermore, we show that $Z$-axis rotations could be realized with the detuning. Due to the reduction of the need for the ancillary qubit for the logical qubit rotation, our proposed approach is advantageous for quantum computation in the NISQ era, where the number of qubits is limited. 
\end{abstract}

\maketitle
\section{Introduction}
\label{sec:intro}
Quantum states 
\YM{are affected by decoherence} under the influence of noise~\cite{nielsen2010quantum,cho2020biggest}. Therefore, maintaining long-time coherence through quantum error correction (QEC) is crucial for performing large-scale quantum computation~\cite{shor1995scheme,laflamme1996perfect,steane1996error}. Additionally, methods have been proposed to enhance the performance of \YM{noisy intermediate-scale quantum \AdMORI{(NISQ)} computation} 
using QEC principles~\cite{mcclean2020decoding,endo2022quantum,tsubouchi2023virtual}.

Recently, QEC methods based on bosonic modes have garnered significant attention ~\cite{terhal2020towards,cai2021bosonic,joshi2021quantum,ma2021quantum,hu2019quantum,rosenblum2018fault,wang2019heisenberg,escher2011general}. In bosonic systems, each mode possesses an infinitely large Hilbert space, which can be leveraged for QEC~\cite{cai2021bosonic}. The primary advantage of this approach over conventional methods is the reduction in the number of required elements. Experimental evidence has been demonstrated that the lifetime of a quantum state is extended after applying error correction in bosonic mode QEC~\cite{ofek2016extending}. Superconducting resonators have emerged as a prominent implementation method for qubits using bosonic modes, where photon loss is the main source of decoherence~\cite{leghtas2013deterministic,ma2021quantum,mizuno2023effect}. The impact of decoherence on these logical qubits has been numerically verified, and theoretical proposals for achieving optimal gate operations have been developed~\cite{mizuno2023effect}.

The method of embedding qubits within the degrees of freedom of a bosonic system to enable error correction is known as a bosonic code. Bosonic codes include the Gottesman-Kitaev-Preskill (GKP) code~\cite{GKP_2001}, the cat code~\cite{Leghtas_2013cat}, and the binomial code~\cite{Marios2016}. Among these \YM{bosonic codes}, the binomial code offers resistance to decoherence mechanisms such as photon loss and phase relaxation if we consider the higher order~\cite{hu2019quantum}. Additionally, operations with binomial codes have the advantage of requiring fewer Fock states compared to other bosonic codes~\cite{hu2019quantum}. However, the implementation of single-qubit operations has been challenging for two main reasons. First, it necessitates the use of pulses optimized with the gradient ascent pulse engineering (GRAPE) method. Second, it requires the use of 
\YM{ancillary} qubits~\cite{khaneja2005optimal,de2011second,mizuno2023effect}.

\YM{Kerr-nonlinear parametric oscillators (KPOs) are regarded as promising platforms for quantum information processing}~\cite{cochrane1999macroscopically,bartolo2016exact,goto2016bifurcation,puri2017engineering,minganti2018spectral}. 
\YM{By strongly parametric driving a Kerr-nonlinear resonator, we induce a bifurcation process known as parametric excitation~\cite{milburn1991quantum,wielinga1993quantum,goto2016bifurcation,puri2017engineering}. For this bifurcation to occur, the Kerr nonlinearity must exceed the photon loss rate}
~\cite{milburn1991quantum,wielinga1993quantum,goto2016bifurcation,puri2017engineering}. 
\YM{This condition is achievable with superconducting quantum circuits and has been confirmed in 
experimental demonstrations}~\cite{wang2019quantum,grimm2020stabilization,yamaji2022spectroscopic}.

\YM{In this paper, we propose a method to perform single-qubit rotations of a logical qubit defined by the binomial code without the need for an ancillary qubit. Instead of relying on an ancillary qubit, we utilize the Kerr nonlinearity of the resonator and the parametric drive. Specifically, we demonstrate that applying two parametric drives with different frequencies simultaneously to a Kerr-nonlinear resonator can implement $X$-axis rotations. Additionally, we propose a technique to achieve $Z$-axis rotations by adjusting a Kerr nonlinearity. 
}

\YM{We perform numerical simulations to evaluate the performance of our quantum error correction method using binomial codes, incorporating the effects of single-photon loss into a master equation. We analyze the dynamics of the binomial code under single-photon loss and find that our quantum error correction method effectively recovers fidelity despite the presence of photon loss during the correction process. Since our single-qubit operations do not require ancillary qubits, our method is particularly well-suited for NISQ (Noisy Intermediate-Scale Quantum) computing, where qubit resources are limited. Furthermore, this method can be implemented using existing experimental devices developed for KPOs.}

\section{Kerr-nonlinear resonator}
In this section, we describe Kerr nonlinear resonators with the parametric drive~\cite{milburn1991quantum,wielinga1993quantum,cochrane1999macroscopically,goto2016bifurcation,puri2017engineering}. 
\YM{Such} a bosonic system \YM{can be} described by the \YM{following} Hamiltonian
\begin{equation}
\label{hpar}
\begin{split}
\hat{H}={}&\chi\hat{N}^2+\omega_{\rm{knr}} \hat{N}
         +{\rm{\textcolor{black}{2}}}p(\hat{a}^{2}+\hat{a}^{\dag 2})\cos \omega 't \\ &+{\rm{\textcolor{black}{2}}}r(\hat{a}+\hat{a}^{\dag})\cos \omega ''t,
\end{split}
\end{equation}
where $\chi$ is a Kerr nonlinearity, $\hat{N}$ is the number operator defined as $\hat{N} = \hat{a}^{\dagger}\hat{a}$, $\omega_{\rm{knr}}$ is the resonance frequency of the Kerr nonlinear resonator, $p$ is the amplitude of the parametric drive, $\omega '$ is the 
frequency \YM{of the parametric drive}, $r$ is the 
amplitude \YM{of the coherent drive}, and $\omega ''$ is the
frequency \YM{of the coherent drive}. 
\YM{
By changing the applied magnetic flux in the SQUID loop~\cite{yamaji2022spectroscopic},
it is possible to tune $\omega_{\rm{knr}}$ and $\chi$, which we assume throughout the paper.}
Moving to a frame rotating at a frequency of $\omega ''$ and applying the rotating wave approximation, we transform the Hamiltonian \YM{as follows:}
\begin{eqnarray}
\hat{H}&\simeq& {}\chi\hat{N}^2+\Delta\hat{N}
         -p(\hat{a}^{2}e^{i\omega t}+\hat{a}^{\dag 2}e^{-i\omega t})\nonumber \\
         &&+r(\hat{a}+\hat{a}^{\dag}),\ \     \label{kponormalh}
\end{eqnarray}
where $\Delta=\omega_{knr}-\omega''$
\YM{is}
the detuning between
\YM{the frequency of the Kerr nonlinear resonator 
and that of the rotating frame}.
Also, $\omega=\omega'- 2\omega''$ is the detuning between the frequency of the parametric drive and twice of that of the coherent drive.

\section{Binomial codes}
\YM{Binomial codes are one of the quantum error correcting codes for bosonic systems. Here, a superposition of the Fock states is utilized, and the weight is the binomial coefficient.}
At higher orders, the binomial code can correct errors 
such as photon loss, 
photon amplification, 
phase relaxation, and their combinations. The lowest-order binomial code is defined as
\begin{align}
\begin{split}
&\ket{0_L} =(\ket{0} +\ket{4})/\sqrt{2},  \\
&\ket{1_L} =\ket{2}.
\end{split}
\end{align}
This code is capable of correcting photon loss, which is described by the noise operator of $\hat{a}$.
\\
\indent 
\YM{Let us consider}
a \YM{single}
logical qubit state $\alpha \ket{0_L}+\beta\ket{1_L}$, where $\alpha$ and $\beta$ are arbitrary complex numbers satisfying the normalization condition $|\alpha|^2+|\beta|^2=1$. In the event of photon loss, the resulting state can be described as follows:
\begin{equation}
    \hat{a}(\alpha \ket{0_L} + \beta \ket{1_L})=\sqrt{2}(
    \alpha \ket{0_E} + \beta \ket{1_E}).
\end{equation}
Here, $\ket{0_E}$ and $\ket{1_E}$ are defined as follows:
\begin{align}
\begin{split}
&\ket{0_E}=\ket{3}, \\
&\ket{1_E}=\ket{1}.
\end{split}
\end{align}
In this case, $\ket{0_L}$ and $\ket{1_L}$ have even photon numbers, while 
$\ket{0_E}$ and $\ket{1_E}$ have odd photon numbers. In the event of photon loss errors, we can recover the original state $\alpha\ket{0_L}+\beta\ket{1_L}$ by performing a parity measurement to determine whether the photon counts are even or odd~\cite{riste2013deterministic}, and then applying the appropriate recovery operation.

However, previous 
\YM{method} required complex pulse operations and 
\YM{ancillary} qubits for performing $X$-axis rotations between $\ket{0_L}$ and $\ket{1_L}$ \cite{Marios2016,hu2019quantum,mizuno2023effect}. In NISQ devices, the use of 
\YM{ancillary} qubits \YM{for the control}
presents a significant challenge.

\section{Rotating operation of a single logical qubit in a Binomial code}
\subsection{Theoretical proposal}
First, we explain the proposed method for $X$-axis rotation of a single logical qubit in a binomial code. 
Although we consider the coherent drive and parametric drive in Eq \eqref{kponormalh}, we do not consider the coherent drive but consider two types of parametric drive with different frequencies.
\YM{In this case, the Hamiltonian is written as follows}
\begin{equation}
\begin{split}
\label{ham}
\hat{H}={}&-4\chi \hat{N}+\chi \hat{N}^2  +p_1(\hat{a}^2 e^{i\omega t}+\hat{a}^{\dagger 2} e^{-i\omega t})   \\ &+p_2(\hat{a}^2 e^{-i\omega t}+\hat{a}^{\dagger 2} e^{i\omega t}).
\end{split}
\end{equation}
where we move to a frame rotating at a frequency of $\omega ''$ and apply the rotating wave approximation.
Here, a two-frequency parametric drive is applied simultaneously, with 
$p_1$ and $p_2$ representing the 
\YM{amplitude} of each parametric drive. The detuning is set to \YM{be} $\Delta\equiv \omega_{knr}-\omega''=-4\chi$. \YM{Let us define the following Hamiltonian:}  
\begin{equation}
    \hat{H}_0=-4\chi \hat{N}+\chi \hat{N}^2.
    \label{hzerohamiltonian}
\end{equation}
\AdMORI{Using} a unitary operator
\begin{equation}
    \hat{U}=e^{i\hat{H}_{0} t},
\end{equation}
\YM{we move} to the rotating frame 
\YM{where we set} $\omega =-4\chi$ in Eq.~\eqref{ham}. Applying the rotating wave approximation, we obtain:
\begin{equation}
\hat{H} \simeq \sqrt{2}p_1(|0\rangle\langle2|+|2\rangle\langle0|)+\sqrt{12}p_2(|2\rangle\langle4|+|4\rangle\langle2|).
\end{equation}
It is worth mentioning that we drop the counter-rotating term here.
We will study the validity of this approximation via numerical simulations later.
Here, \YM{by} setting $p_1=\sqrt{6}p_2$, we obtain
\begin{eqnarray}
\hat{H} &\simeq& \sqrt{12}p_2\Big{(}\left(|0\rangle+|4\rangle\right)\langle2|+|2\rangle\left(\langle0|+\langle4|\right)\Big{)} \nonumber \\
&=&\sqrt{24}p_2 (|0_L\rangle\langle1_L|+|1_L\rangle\langle0_L|).
\end{eqnarray}
\YM{So, by} applying two parametric drives with different amplitudes and frequencies, \YM{we} can induce Rabi oscillations between $|0_L\rangle$ and $|1_L\rangle$. Therefore, by adjusting $p_2$, the $X$-axis rotation operation of a single logical qubit in the binomial code can be performed 
\YM{with} an arbitrary angle.

Next, we explain the method 
\YM{to perform} $Z$-axis rotation of a single logical qubit on a Binomial code. 
\YM{Let us}
consider the subspace spanned by $\ket{0}$, $\ket{2}$ and $\ket{4}$. 
\YM{Also, we} set $p_1=p_2=0$. In this case, Eq. (6) \YM{is}
simplified to:
\begin{equation}
\begin{split}
\label{hamZ}
\hat{H}=-4\chi |2\rangle \langle 2| .
\end{split}
\end{equation}
By evolving the system under this Hamiltonian, we obtain $e^{i4\chi t |2\rangle \langle 2|}(\alpha \ket{0_L} +\beta \ket{1_L})=\alpha \ket{0_L} +e^{i4\chi t}\beta \ket{1_L}$. Thus, an arbitrary relative phase can be applied between 
$\ket{0_L}$ and $\ket{1_L}$. This means that a $Z$-axis rotation of a single logical qubit in a binomial code can be performed by adjusting a Kerr nonlinearity. 
 
\subsection{Numerical calculations}
\begin{figure}[h!]
    \centering
    \includegraphics[width = 8.5cm]{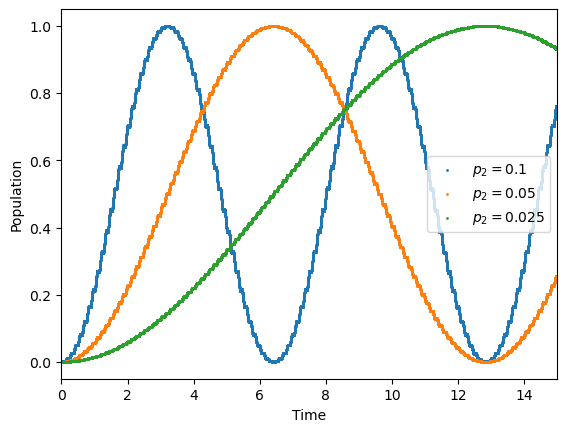}
    \caption{
    \YM{Rabi oscillation between $\ket{1_L}$ and $\ket{0_L}$ with our method.}
    The horizontal axis represents time $t$, and the vertical axis represents the population of $\ket{1_L}$. The parameters are set to $\chi = 6$ and $p_1 = \sqrt{6}p_2$. As $p_2$ increases, high-frequency oscillations begin to be observed \YM{due to the breakdown of rotating wave approximation}.
    }
    \label{kaiten_2}
\end{figure}
\begin{figure}[h!]
    \centering
    \includegraphics[width = 8.5cm]{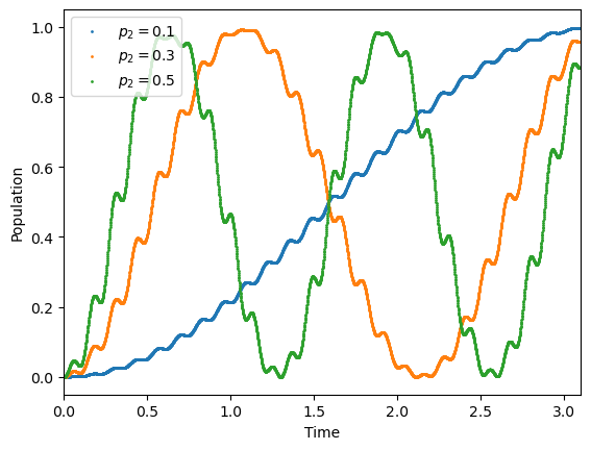}
    \caption{
    \YM{Rabi oscillation between $\ket{1_L}$ and $\ket{0_L}$ with our method.}
   The horizontal axis represents time $t$, and the vertical axis represents the population of $\ket{1_L}$. The parameters are set 
   \YM{as} $\chi = 6$ and $p_1 = \sqrt{6}p_2$. As $p_2$ increases, high-frequency oscillations, resulting from the breakdown of the rotating wave approximation, 
   \YM{become more apparent}.
    }
    \label{kaiten}
\end{figure}
\YM{We quantify the performance of our method by numerical simulations.}
\YM{More specifically,}
we present numerical results of Rabi oscillations between $\ket{0_L}$ and $\ket{1_L}$, which are expected to be influenced by the rotating wave approximation. We adopt the Hamiltonian 
\YM{described in}
Eq.~\eqref{ham} and set the parameters 
\YM{as} $\omega=-4\chi$ and $p_1=\sqrt{6}p_2$
 . The initial state is $\ket{0_L}$, and the population of $\ket{1_L}$ during \YM{the}
 time evolution is plotted in Figure \ref{kaiten_2}.
 Figure \ref{kaiten} shows the Rabi oscillations between $|0_L\rangle$ and $|1_L\rangle$ where we vary the values of $p_1$ and $p_2$.
From Figs. \ref{kaiten_2} and \ref{kaiten}, we observe that as the 
\YM{amplitude} of the parametric drive increases, the rotating wave approximation breaks down, resulting in high-frequency oscillations. Therefore, to 
\YM{perform high-fidelity}
Rabi oscillations, the parametric drive must be applied weakly.
\YM{This is a stark contrast to the conventional KPO, where the nonlinear resonator is strongly driven by the parametric drive~\cite{milburn1991quantum,wielinga1993quantum,cochrane1999macroscopically,goto2016bifurcation,puri2017engineering}.}

\begin{figure}[h!]
    \centering
    \includegraphics[width = 8.5cm]{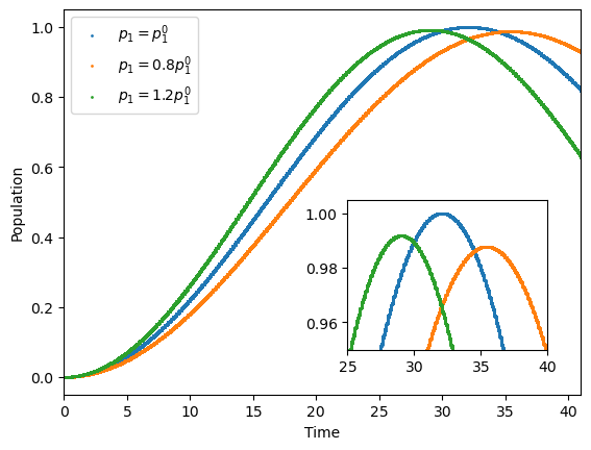}
    \caption{
    Rabi oscillations between $\ket{0_L}$ and $\ket{1_L}$ are shown for varying values of $p_1$ from $p_{1}^{0} \equiv \sqrt{6} p_2$. The parameters are 
    $\chi = 6$ and $p_2 = 0.01$. When 
    \quad$p_1=p_{1}^{0}$, \quad$p_1=0.8p_{1}^{0}$, and \quad$p_1=1.2p_{1}^{0}$, the maximum populations of $\ket{1_L}$ are 0.99999, 0.98780, and 0.99180, respectively.
    }
    \label{hi}
\end{figure}

In addition, the condition $p_1=\sqrt{6}p_2$
  is necessary for our method to induce Rabi oscillations between 
$\ket{0_L}$ and $\ket{1_L}$. However, it may not always be possible to adjust this parameter precisely in actual experiments. Therefore, we evaluate the accuracy of the Rabi oscillation 
\YM{for} 
$p_1\neq\sqrt{6}p_2$
  through numerical calculations.
From Fig.~\ref{hi}, 
\YM{for} $p_1=\sqrt{6}p_2(\equiv p^{0}_1)$, the population of $\ket{1_L}$ is nearly 1. However, 
\YM{for} 
$p_1=0.8p^{0}_1$ (or $p_1=1.2p^{0}_1$), the population of 
$\ket{1_L}$ is reduced to approximately 0.9878 (or 0.9918). This indicates that the fidelity of the $X$-axis rotation can exceed 0.987 even with a 20\% deviation from the condition $p_1=\sqrt{6}p_2$.

\section{Quantum error correction}
\subsection{Theoretical proposal}
We describe the procedure for correcting quantum errors due to single-photon loss in our 
\YM{method}. As mentioned above, when a single-photon loss occurs in the state $\alpha \ket{0_L} + \beta \ket{1_L}$, the resulting state becomes
\begin{align}
\hat{a}(\alpha \ket{0_L} + \beta \ket{1_L}) = \sqrt{2}(\alpha \ket{3} + \beta \ket{1}).
\end{align}
By performing a parity measurement, which determines whether the photon number is odd or even, the occurrence of single-photon loss can be detected \cite{sun2014tracking,hu2019quantum}. 
This type of parity measurement requires an 
\YM{ancillary} qubit. Specifically, in the rotating frame, the Hamiltonian \YM{to describe the coupling with the ancillary qubit}
is given by $\hat{H} = \frac{g^2}{\Delta'}\hat{a}^{\dagger} \hat{a}\hat{\sigma}_z$ (dispersive interaction) \YM{where we set $\chi=0$ for the resonator}. 
Here, $\Delta'$ represents the detuning between the 
\YM{ancillary} qubit and the resonator, $g$ is the coupling constant between the 
\YM{ancillary} qubit and the resonator, and $\hat{\sigma}_z$ is the Pauli matrix acting on the 
\YM{ancillary} qubit.
The resonator is coupled to the 
\YM{ancillary} qubit, and a measurement is performed on the 
\YM{ancillary} qubit \YM{for the parity measurement}
\cite{sun2014tracking,hu2019quantum}.  

Our method, while necessitating 
\YM{ancillary} qubit for parity measurement, 
does not require 
\YM{the ancillary qubit} for \YM{the recovery operation during the quantum error correction.}
This is a significant advantage over conventional approaches \cite{Marios2016,hu2019quantum,mizuno2023effect}. 
\\

\YM{Let us explain our method for the recovery operation during the quantum error correction. Here,
starting from $(\alpha \ket{3} + \beta \ket{1})$, we aim to obtain $\alpha \ket{0_L} + \beta \ket{1_L}$.}

\subsubsection{Transition from $\ket{3}$ to $\ket{2}$}

Using the coherent drive, we \YM{perform Rabi oscillation between}
$|3\rangle$ and $|2\rangle$.
The Hamiltonian is described as
\begin{align}
\hat{H}_{32}=-4\chi \hat{N}+\chi \hat{N}^2+\lambda_{32}(\hat{a}e^{i\chi t}+\hat{a}^{\dagger}e^{-i\chi t})\label{Hamil_32},
\end{align}
where the parameter $\lambda_{32}$ represents the 
\YM{amplitude of the coherent drive}. The \YM{time} duration $t_{32}$ for applying the $\pi$ pulse 
\YM{between}
$\ket{3}$ 
\YM{and} $\ket{2}$ can be determined via the \YM{so called}
Rabi frequency $\omega_{32}$, which depends on $\lambda_{32}$. Analytically, the relationship is given by:
\begin{align}
\omega_{32}=2\sqrt{3}\lambda_{32},\quad t_{32}=\pi/\omega_{32}\label{constraint_32}.
\end{align}
Thus, we adjust $\lambda_{32}$ and $t_{32}$ to satisfy this condition.

By applying this procedure, the state $\ket{\psi_{32}}=\alpha \ket{2} + \beta e^{i\theta_{1}}\ket{1}$ is obtained, where $\theta_{1}$ represents the relative phase between states $\ket{1}$ and $\ket{2}$.

\subsubsection{Transition from $\ket{2}$ to $\ket{0_L}$}
Using parametric drive, we induce Rabi oscillations between $\ket{2}$ and $\ket{0_{L}}$. The Hamiltonian is expressed as:
\begin{align}
    \hat{H}_{204}=-4&\chi\hat{N}+\chi\hat{N}^2+p_1(\hat{a}^2e^{-i4\chi t}+\hat{a}^{\dagger 2}e^{+i4\chi t})\nonumber\\
    &+p_2(\hat{a}^2e^{i4\chi t}+\hat{a}^{\dagger 2}e^{-i4\chi t})\label{Hamil_204}
\end{align}
The conditions for the Rabi frequency $\omega_{204}$, $\pi$ pulse duration $t_{204}$, and $p_1$ and $p_{2}$ are given by:
\begin{align}
    p_{1}=\sqrt{6}p_{2},\ \omega_{204}=4 p_{1},\ t_{204}=\pi/\omega_{204}\label{constraint_204}.
\end{align}

\subsubsection{Transition from $\ket{1}$ to $\ket{1_L}$}

\YM{Using the coherent drive, we perform Rabi oscillation between}
$\ket{1}$ and $\ket{2}$. The Hamiltonian is described as:
\begin{align}
    H_{12}=-4\chi \hat{N}+\chi \hat{N}^2+\lambda_{12}(\hat{a}e^{-i\chi t}+\hat{a}^{\dagger}e^{i\chi t})\label{Hamil_12}.
\end{align}
In this step, the 
\YM{conditions of}
the Rabi frequency $\omega_{12}$, $\pi$ pulse duration $t_{12}$, and the pulse intensity $\lambda_{12}$ are given by:
\begin{align}
    \omega_{12}=2\sqrt{2}\lambda_{12},\ t_{12}=\pi/\omega_{12}\label{constraint_12}.
\end{align}
\YM{By performing the $\pi$ pulse,}
we obtain $\ket{\psi_{12}}=\alpha \ket{0_L} + \beta e^{i\theta_3}\ket{1_L}$. Here, $\theta_{3}$ represents the relative phase between states $\ket{0_L}$ and $\ket{1_L}$.

\subsubsection{Adjustment of the relative phase}

\YM{We perform the $Z$-axis rotation operation on a single logical qubit in the binomial code.}
We set the Hamiltonian as \YM{follows}
\begin{align}
 \hat{H}=-4\chi\hat{N}+\chi\hat{N}^2.\label{Hamil_phadj}
\end{align}
This allows us to restore \YM{the state of} $\ket{\psi}=\alpha \ket{0_L} + \beta \ket{1_L}$.

Thus, \YM{we can perform the recovery operation of the quantum error correction in the binomial code without an ancillary qubit.}

\subsection{Numerical calculations}
\YM{We perform numerical simulations to evaluate the performance of the quantum error correction of our method.}
We choose the initial state \YM{as}
$\ket{\psi}=\frac{1}{\sqrt{2}}\ket{0_L} + \frac{1}{\sqrt{2}} \ket{1_L}$
and incorporate the effects of decoherence using 
\YM{the Gorini-Kossakowski-Sudarshan-Lindblad (GKSL)}
master equation \cite{gorini1976completely,lindblad1976generators} \YM{as follows:}
\begin{align}
\dot{\rho}=-i[\hat{H},\rho]+\hat{L}\rho\hat{L}^{\dagger}-\frac{1}{2}\{\hat{L}^{\dagger}\hat{L},\rho\}.
\end{align} 
We use Eq.~\eqref{hzerohamiltonian} with $\chi=0$ as the Hamiltonian, and the Lindblad operator representing single-photon loss is given by
\begin{equation}
\hat{L}=\sqrt{\gamma}\hat{a},
\end{equation}
with $\gamma$ representing the decoherence rate. 
\YM{Throughout the paper,}
we set $\gamma=0.001$.
By solving this master equation from 
$t=0$ to $t=t_{\rm error}$, we obtain the density matrix $\rho(t_{\rm error})$ where $t_{\rm error}$ is the exposure time to decoherence. We then perform a photon number parity measurement on the resultant state, \YM{which is assumed to be perfect for our numerical simulations}. Assuming an error is detected, we simulate the error correction procedure for such cases. 
Specifically, using the parity projection operator: 
\YM{
\begin{eqnarray}
    \hat{P}_{\rm odd}=\sum _{n=0}^{\infty}|2n+1\rangle \langle 2n+1|.
\end{eqnarray}
}
This operator projects \YM{the state into a subspace spanned by odd-number Fock states.}
The density matrix immediately after the error detection is given by
\begin{equation}
{\rho}' = \frac{\hat{P}_{\rm odd}\rho(t_{\rm error})\hat{P}_{\rm odd}}{\mathrm{Tr}[\hat{P}_{\rm odd}\rho(t_{\rm error})\hat{P}_{\rm odd}]}.
\end{equation}
We then apply our 
\YM{recovery process for quantum error correction}
to this 
${\rho}'$. 
 \textcolor{black}{In the numerical simulation, we consider the effects of decoherence (single-photon loss) during the execution of the recovery process for the quantum error correction and the imperfections of the gate operations.}
 \YM{To quantify the performance of our proposal, we calculate}
 the fidelity between the density matrix after quantum error correction and the 
 state 
$\ket{\psi}$, \YM{which we aim to restore after the quantum error correction}.

\begin{figure*}[tb]
    \centering
    \includegraphics[bb=0 0 1000 300,width=18cm]{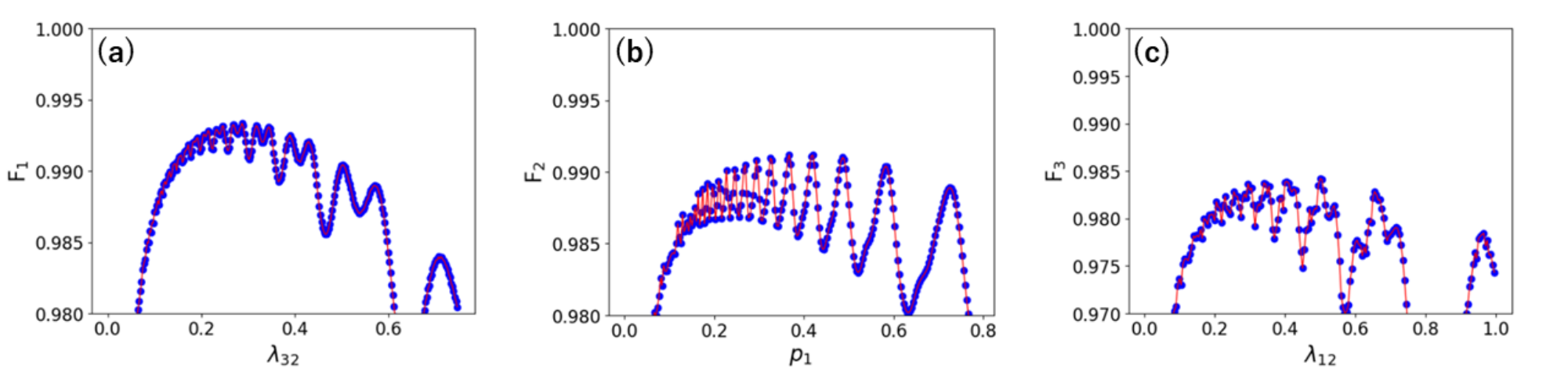}
    \caption{Optimization of 
    \AdMORI{the amplitudes} for each step: (a) Transition from $\ket{3}$ to $\ket{2}$, (b) Transition from $\ket{2}$ to $\ket{0_L}$, and (c) Transition from $\ket{1}$ to $\ket{2}$. The 
    \YM{horizontal axis} represents the 
    \YM{amplitude of the external fields such as} $\lambda_{32}$, $p_1$, and $\lambda_{12}$ for each step. The 
    \YM{vertical axis} denotes the fidelity of the states 
    \YM{obtained from}
    the parameters $\lambda_{32}$, $p_1$, and $\lambda_{12}$.
    In the first step, the fidelity is maximized at $\lambda_{32} = 0.2875$, yielding a fidelity of 0.99333. In the second step, the fidelity peaks at $p_1 = 0.3675$, resulting in a fidelity of 0.99120. In the third step, the maximum fidelity is attained at $\lambda_{12} = 0.500$, with a corresponding fidelity of 0.98420.}
    \label{fig:bystep_fid}
\end{figure*}

Our proposed method employs three types of $\pi$ pulses for error correction. 
To mitigate the effects of decoherence, the duration of the pulse should be minimized, \YM{and so we need to increase the Rabi frequency}. However, 
\YM{the increase of the Rabi frequency} can disrupt the rotating wave approximation and potentially deteriorate the fidelity due to the \YM{unwanted transitions} to non-target levels. Thus, identifying the optimal pulse strength is crucial in this trade-off scenario. 

We optimize the parameters $\lambda_{32}$, $p_{1}$, $\lambda_{12}$ to achieve the highest fidelity. More specifically, we use the following strategy. 

\begin{figure}[h!]
    \centering
    \includegraphics[width=8.5cm]{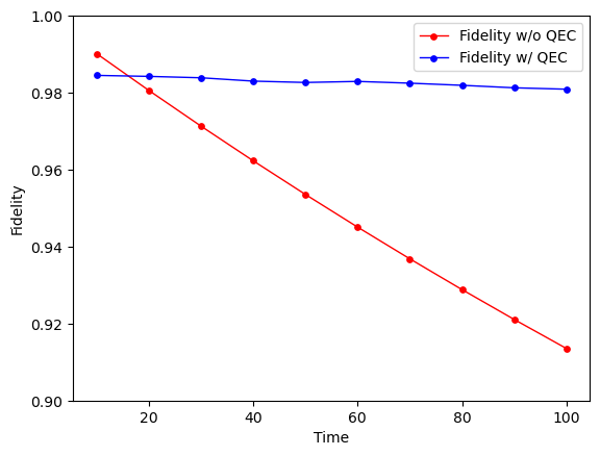}
    \caption{ 
    \YM{Comparison between the fidelity with and without the quantum error correction.}
    The horizontal axis represents the exposure time $t_{\rm error}$, and the vertical axis represents the fidelity. ``Fidelity w/o QEC'' indicates the fidelity before correction, while ``Fidelity w/ QEC'' indicates the fidelity after correction. The parameters $\chi = 6$ and $\gamma = 0.001$ are used.}
    \label{fid_graph}
\end{figure}
To determine the optimal value of $\lambda_{32}$ for the state $\rho_{32}$ obtained after the first step, we numerically calculate the fidelity $F_1 = \max_{\theta} \langle \psi_{32} |\hat{U}^{\dagger}_{\theta} \rho_{32} \hat{U}_{\theta} |\psi_{32} \rangle$. Here, the operator $\hat{U}_{\theta} = e^{-i(-4\hat{N} + \hat{N}^2)\theta}$ is employed for phase correction. Subsequently, using the state $\rho_{32}$ obtained with this optimal $\lambda_{32}$, the second step is performed. For the state $\rho_{204}$ obtained from the second step, 
\YM{we calculate}
the fidelity $F_2 = \max_{\theta} \langle \psi_{204} |\hat{U}_{\theta}^{\dagger} \rho_{204} \hat{U}_{\theta} |\psi_{204} \rangle$
\YM{and choose $p_1$ and $p_2$ to maximize this fidelity, which we will use for the third step. }
For the state $\rho_{12}$ obtained from the third step, the fidelity $F_3 = \max_{\theta} \langle \psi_{12} |\hat{U}_{\theta}^{\dagger} \rho_{12} \hat{U}_{\theta} |\psi_{12} \rangle$ is evaluated, and \YM{we choose the value of $\lambda_{12}$ to maximize the fidelity, which we will use for the final step.}
Finally, using the state $\rho_{12}$, 
the final fidelity $F_{\rm{fin}} = \max_{\theta} \langle \psi |\hat{U}_{\theta}^{\dagger} \rho_{12} \hat{U}_{\theta} |\psi \rangle$ is obtained. The relationship between \YM{the parameters} ($\lambda_{32}$, $p_1$, and $\lambda_{12}$) and the fidelity of the obtained state is illustrated in Fig. \ref{fig:bystep_fid}.

Consider the fidelity $F_{\rm{ini}} = \langle \psi | \rho(t_{\rm error}) | \psi \rangle$ between the state $\rho(t_{\rm error})$ prior to error correction and the ideal state $\ket{\psi}$. 
\YM{We plot the this initial fidelity $F_{\rm{ini}}$ and the final fidelity $F_{\rm{fin}}$ against the time duration $t_{\rm error}$ in Fig. \ref{fid_graph}.
As we increase $t_{\rm error}$, the decoherence becomes more relevant, and both fidelities decrease.
However, the gradient for $t_{\rm error}$ with quantum error correction is 
smaller 
compared to the case without the quantum error correction.
This means that our method effectively restores the state for longer $t_{\rm error}$. On the other hand, for shorter $t_{\rm error}$, our method does not help to restore the state because of the accumulation of errors during the recovery process.
}

\section{Conclusion}
\YM{Here, we propose a method to perform single-qubit rotations for the binomial code without ancillary qubits.}
To implement the $X$-axis rotations, we simultaneously apply two parametric drives with distinct frequencies to a 
\YM{Kerr nonlinear} resonator. Additionally, we 
\YM{can implement}
$Z$-axis rotations through detuning. 
Moreover, \YM{we perform numerical simulations to quantify the performance of our proposed method and confirm that our method helps to restore the state.}
This approach holds particular promise for NISQ devices by mitigating the reliance on ancillary qubit \YM{when we perform single-qubit rotations of the logical qubit on the binomial code}.

\begin{acknowledgements}
We would like to thank Suguru Endo for fruitful discussions. We also would like to thank the developers of QuTiP~\cite{Qutip}, which was used for our numerical simulations.
This paper is partly
based on results obtained from a project, JPNP16007,
commissioned by the New Energy and Industrial Technology Development Organization (NEDO), Japan.
This work was supported by
JST Moonshot (Grant Number JPMJMS226C). Y. Matsuzaki is supported by JSPS KAKENHI (Grant Number
23H04390). This work was also supported by CREST
(JPMJCR23I5), JST.
\end{acknowledgements}




\bibliography{ref}

\end{document}